Body Diagonal Diffusion Couple Method for Estimation of Tracer Diffusion Coefficients in a Multi-Principal Element Alloy

Anuj Dash, and Aloke Paul*
Department of Materials Engineering, Indian Institute of Science, Bangalore 560012
Corresponding author: aloke@iisc.ac.in

**Abstract**

The estimation of $(n-1)^2$ interdiffusion coefficients in an n component system requires (n-1) diffusion paths to intersect or pass closely in the (n-1) dimensional space according to the body diagonal diffusion couple method. These interdiffusion coefficients are related to n(n-1) intrinsic (or n tracer diffusion coefficients), which cannot be estimated easily following the Kirkendall marker experiment in a multicomponent system despite their importance for understanding the atomic mechanism of diffusion and the physico-mechanical properties of materials. In this study, the estimation of tracer diffusion coefficients from only two diffusion profiles following the concept of the body diagonal diffusion couple method in a multicomponent system is demonstrated. Subsequently, one can estimate the intrinsic and interdiffusion coefficients. This reduces the overall effort up to a great extent since it needs only two instead of (n-1) diffusion profiles irrespective of the number of components, with an additional benefit of enabling the estimation of all types of diffusion coefficients. The available tracer diffusion coefficients estimated following the radiotracer method are compared to the data estimated in this study following this method. This method can also be extended to the systems in which the radiotracer method is not feasible.

## Introduction

It is extremely challenging (or rather impossible) to intersect ($n$-1) diffusion paths strictly at one composition for $n$>3 in the multicomponent space for estimating the $(n-1)^2$ interdiffusion coefficients, which was an unsolved problem over the last many decades [1]. A few alternate diffusion couple methods have been established recently to circumvent this problem [2-5]. The body diagonal diffusion couple method is one such method proposed by Morral [5] for the estimation of the interdiffusion coefficients in which diffusion couples are designed in a small composition range of almost constant diffusivities such that the diffusion paths pass closely even if they do not intersect exactly at a single composition [6]. This facilitates the estimation of the interdiffusion coefficients from the calculated interdiffusion flux and concentration gradients from composition profiles at the closest compositions. This brings a very positive development for analyzing the interdiffusion in a multicomponent system. However, still, this is not enough since these interdiffusion coefficients are related to $n(n-1)$ unknown intrinsic diffusion coefficients or $n$ tracer diffusion coefficients in correlation with the $n(n-1)$ thermodynamic factors [7, 8]. The knowledge of intrinsic and tracer diffusion coefficients is essential for understanding the fundamental atomic mechanism of transport or numerical and simulation methods dealing with various physical and mechanical properties of materials. These parameters cannot be estimated following the Kirkendall marker experiment in a system with more than two components since it is almost impossible to predict the end-member compositions of all the diffusion couples such that this plane is located at the composition of the intersection or at the closest compositions when the body diagonal diffusion couple method is followed. In binary systems, the estimation of these parameters helped to understand the atomic mechanism of diffusion in various systems and to develop a physico-chemical approach in correlation to the microstructural evolution [9]. For example, these diffusion parameters explained a very strong role of the thermodynamic driving force in the $Nb_3Sn$ system [10]. Even the growth of this phase mainly because of grain boundary but not the lattice diffusion could be identified by comparing the diffusion rates of the components [11]. Sometimes, strong contributions from both lattice and grain boundary diffusion could be identified for the growth of the product phases [12, 13]. In β-Ni(Pt)Al point defects play a dominant role over the thermodynamic driving forces, which could be understood following such analysis [14, 15]. Therefore, a method for the estimation of tracer and intrinsic diffusion coefficients is very important for multicomponent diffusion profiles.

Kirkaldy and Lane [16] proposed the estimation of the tracer diffusion coefficients from the estimated interdiffusion coefficients at the cross of diffusion paths in a ternary system utilizing the thermodynamic parameters. Following, one can calculate the intrinsic diffusion coefficients. This study connects the Kirkaldy-Lane method to the body diagonal diffusion couples in NiCoFeCr MPE alloy for the estimation of all the diffusion parameters. Instead of estimating the tracer diffusion coefficients from the estimated interdiffusion coefficients which need (n-1) diffusion paths to pass closely (or intersect) as proposed by Kirkaldy and Lane [16], we have shown that the estimation of tracer diffusion coefficients directly from the interdiffusion flux needs only two diffusion profiles passing closely irrespective of the number of components. From the estimated tracer diffusion coefficients one can then estimate the intrinsic and even the interdiffusion coefficients, which reduces the effort up to great extent in multicomponent systems.

## 2. Experimental method

The alloys used for preparing diffusion couples were arc-melted in an argon atmosphere with pure elements (99.95 - 99.99 wt.%). The alloy buttons were then annealed for homogenization at 1200 ± 5°C for 50 hrs in a high vacuum ($\sim 10^{-4}$ Pa). Following, compositions were measured at multiple spots randomly in an EPMA (Electron Microprobe Analyzer) to find a variation within ±0.15 at. % from the average compositions. Approximately 1.5 mm thick slices were cut from the button using an electro-discharge machine and prepared metallographically for smooth and plane parallel surfaces. Two end-member alloys were then diffusion coupled in a special fixture and diffusion annealed at 1200 ± 5°C for 50 hrs. After completion of diffusion annealing, the samples were quenched in water, cross-sectioned, and prepared metallographically for composition profile measurements in EPMA.

## 3. Results and discussion

The end member compositions of the body diagonal diffusion couples, as listed in Table 1, are decided following the concept proposed by Morral [5]. The equiatomic composition was chosen as the body center composition close to which all the diffusion paths may pass (if do not intersect). The composition range (edge length of the cube) of three components is considered as $2\Delta Co = 2\Delta Fe = 2\Delta Cr = 10\ at.\%$ along the $x, y$ and $z$ axes. The composition of the fourth component Ni is therefore calculated from $N_{Ni} = 100 - N_{Co} - N_{Fe} - N_{Cr}$, when $N_i$ is expressed in atomic percent. One may go through Ref. [5] for details on designing alloys for body diagonal diffusion couples. To demonstrate the calculation of both tracer and intrinsic diffusion coefficients, three body diagonal diffusion couples were prepared, although as explained in the article, only two diffusion profiles are required for the estimation of the tracer diffusion coefficients. It helps to demonstrate the consistency of the calculation of tracer diffusion coefficients considering two profiles at a time from three sets of solutions.

Since the diffusion paths do not intersect exactly, we need to find the closest compositions at which the values of interdiffusion fluxes and composition gradients should be considered for the estimation of the diffusion coefficients. This is a crucial step to keep the estimation error as small as possible. The closest compositions can be found by calculating the distance between several points (compositions) on different diffusion paths after identifying the composition range in which the profiles pass close to each other. The composition range of close distances is first identified by examining the 3D tetrahedral plot in Origin software and then studying the diffusion paths from different angular perspectives. Following, the shortest distance is calculated by first converting the four-component barycentric tetrahedral coordinate system to a three-dimensional Cartesian $(x, y, z)$ coordinate system to calculate the Pythagorean distance in Euclidean space. The barycentric representation of the composition space with a tetrahedron having pure Ni, Co, Fe, Cr at the vertices can be redrawn as an equivalent tetrahedron in the cartesian coordinates as shown in Fig. 2 [17].

A composition point in the four-component barycentric space ($N_{Ni}$, $N_{Co}$, $N_{Fe}$, $N_{Cr}$) can be converted into its equivalent cartesian coordinates through geometrical manipulation as explained by Shimura et al. [17] using:

$$x = \frac{N_{Fe} + 1 - N_{Co}}{2} \qquad (1a)$$

$$y = \frac{\sqrt[2]{3}}{2} \times N_{Ni} + \frac{\sqrt[2]{3}}{6} \times N_{Cr} \qquad (1b)$$

$$z = \frac{\sqrt[2]{6}}{3} \times N_{Cr} \qquad (1c)$$

Then the distance between points of different profiles can be obtained using the simple square root formula

$$Distance = \sqrt[2]{(x_1 - x_2)^2 + (y_1 - y_2)^2 + (z_1 - z_2)^2} \qquad (2)$$

The smallest distance obtained after calculating the distances between several points (compositions) on the two diffusion paths in the close pass composition range, identifies the closest points (compositions) between the two diffusion paths. Once the two closest points are identified, the mean point (composition) can be found from

$$N\ (x, y, z) = \frac{1}{2}\sum_{i=1}^{2}(x_i, y_i\ , z_i) \tag{3}$$

As the calculation of tracer diffusion coefficients is done from two diffusion profiles only, the distance of the closest points on the two diffusion paths from a mean point, therefore, can be calculated with the help of

$$Distance\ from\ mean\ = \frac{1}{2}\left(\sqrt[2]{(x_1 - x_2)^2 + (y_1 - y_2)^2 + (z_1 - z_2)^2}\right) \tag{4}$$

The mean position can be stated as the approximate composition at which the tracer diffusion coefficients are estimated. The closest compositions between different diffusion couples are listed in Table 3. The distance to mean between different diffusion paths from each other considering their closest points (very near the equiatomic composition) is found to be in the range of $0.03 - 0.11$ atomic percent.

The relation between interdiffusion flux ($\tilde{J}_i$) of component *i* and interdiffusion coefficients ($\widetilde{D}_{ij}$) in a *n* component system for constant molar volume ($V_m$) is expressed as [8]

$$\tilde{J}_i = -\sum_{j=1}^{n-1}\widetilde{D}_{ij}^{n}\frac{\partial C_j}{\partial x} = -\frac{1}{V_m}\sum_{j=1}^{n-1}\widetilde{D}_{ij}^{n}\frac{\partial N_j}{\partial x} \tag{5a}$$

$$\sum_{i=1}^{n}\tilde{J}_i = 0 \tag{5b}$$

$C_j = \frac{N_j}{V_m}$ (mol/m$^3$) is the concentration of component *j,* where $N_j$ is the composition (mole fraction) and $V_m$ (m$^3$/mol) is the molar volume. $\widetilde{D}_{ii}^{n}$ is the main interdiffusion coefficient of component *i* related to the concentration gradient of the same component and $\widetilde{D}_{ij}^{n}$ is the cross interdiffusion coefficient of component *i*, which is related to the concentration gradient of another component *j*. Composition of component $n$ is considered as the dependent. The interdiffusion flux of component *i* can be calculated directly from the diffusion profile of component *i* using [8]

$$V_m\tilde{J}_i = -\frac{N_i^{+}-N_i^{-}}{2t}\left[(1 - Y_i^{*})\int_{x^{-\infty}}^{x^{*}} Y_i dx + Y_i^{*}\int_{x^{*}}^{x^{+\infty}}(1 - Y_i)dx\right] \tag{6}$$

Where $Y_i = \frac{N_i - N_i^-}{N_i^+ - N_i^-}$ is the composition normalized variable and *t* is the diffusion annealing time.

In a four-component NiCoFeCr system, considering Cr (component 4) as the dependent variable the interdiffusion fluxes of components Ni (component 1), Co (component 2) and Fe (component 3), from Eq. 1 can be expressed as:

$$V_m \tilde{J}_1 = -\tilde{D}_{11}^4 \frac{dN_1}{dx} - \tilde{D}_{12}^4 \frac{dN_2}{dx} - \tilde{D}_{13}^4 \frac{dN_3}{dx} \quad (7a)$$

$$V_m \tilde{J}_2 = -\tilde{D}_{21}^4 \frac{dN_1}{dx} - \tilde{D}_{22}^4 \frac{dN_2}{dx} - \tilde{D}_{23}^4 \frac{dN_3}{dx} \quad (7b)$$

$$V_m \tilde{J}_3 = -\tilde{D}_{31}^4 \frac{dN_1}{dx} - \tilde{D}_{32}^4 \frac{dN_2}{dx} - \tilde{D}_{33}^4 \frac{dN_3}{dx} \quad (7c)$$

These interdiffusion coefficients are related to the unknown intrinsic diffusion coefficients ($D_{ij}^n$) by [8]

$$\tilde{D}_{ij}^n = D_{ij}^n - N_i \sum_{k=1}^{n} D_{kj}^n \quad (8a)$$

In a four-component system considering component 4 as the dependent variable, Eq. 8a can be expressed as

$$\tilde{D}_{ij}^4 = D_{ij}^4 - N_i\left(D_{1j}^4 + D_{2j}^4 + D_{3j}^4 + D_{4j}^4\right) \quad (8b)$$

Therefore, nine interdiffusion coefficients are related to total twelve intrinsic diffusion coefficients and each interdiffusion coefficient is related to four unknown intrinsic diffusion coefficients. The intrinsic coefficients are important parameters to understand interactions between the components and fundamental atomic mechanism of diffusion. But unlike the interdiffusion coefficients that directly relate to the nature of the diffusion (composition) profiles, intrinsic coefficients cannot be calculated from the composition profiles alone. These cannot be estimated following the Kirkendall marker experiment in a system with more than two components since it is almost impossible to find the marker plane position at the composition of intersection or at the closest compositions when the body diagonal diffusion couple method is followed in all the (*n*-1) diffusion couples [8].

Kirkaldy [16] and Le Claire [19] proposed the relation between the intrinsic and tracer diffusion coefficients ($D_i^*$) as

$$D_{ij}^{n} = \frac{N_i}{N_j} D_i^{*} \phi_{ij}^{n} \quad (9)$$

where $\phi_{ij}^{n}$ is the thermodynamic factor expressed as $\phi_{ij}^{n} = \frac{N_j}{RT}\left(\frac{\partial \mu_i}{\partial N_j} - \frac{\partial \mu_i}{\partial N_n}\right)$. $\mu_i$ is the chemical potential of component *i*, R is the gas constant and T is the temperature in K.

As already explained the intrinsic diffusion coefficients cannot be estimated following the Kirkendall marker experiment in a diffusion couple in which more than two components develop the diffusion profiles. Therefore, it is not possible to estimate the tracer diffusion coefficients as well utilizing the thermodynamic parameters because of unknown intrinsic diffusion coefficients. The Kirkendall marker experiment to estimate these parameters is only feasible in binary and pseudo-binary diffusion couples [4, 8]. Replacing Eq. 9 in Eq. 8a [16], the interdiffusion coefficients can be expressed with tracer diffusion coefficients and thermodynamic factors by

$$\widetilde{D}_{ij}^{n} = \frac{N_i}{N_j} D_i^{*} \phi_{ij}^{n} - N_i \sum_{k=1}^{n} \frac{N_k}{N_j} D_k^{*} \phi_{kj}^{n} \quad (10a)$$

In a four-component system considering component 4 as the dependent variable, Eq. 10a can be expressed as

$$\widetilde{D}_{ij}^{4} = \frac{N_i}{N_j} D_i^{*} \phi_{ij}^{4} - N_i \sum_{k=1}^{4} \frac{N_k}{N_j} D_k^{*} \phi_{kj}^{4} \quad (10b)$$

It can be realized that each interdiffusion coefficient is now expressed with *n* tracer diffusion coefficients in an *n* component system, i.e., four tracer diffusion coefficients in a four-component system. Replacing Equation 10 in Equation 5a, the interdiffusion flux can be directly related to tracer diffusion coefficients, thermodynamic factors and composition gradients by

$$\tilde{J}_i = -\frac{1}{V_m} \sum_{j=1}^{n-1} \left[\frac{N_i}{N_j} D_i^{*} \phi_{ij}^{n} - N_i \sum_{k=1}^{n} \frac{N_k}{N_j} D_k^{*} \phi_{kj}^{n}\right] \frac{\partial N_j}{\partial x} \quad (11)$$

It further means that (*n*-1) independent interdiffusion fluxes of a single diffusion path are related to *n* tracer diffusion coefficients (refer Eq. 7 and 11). We need to solve *n* number of equations for estimation of the *n* tracer diffusion coefficients. Therefore, we need only two diffusion paths to intersect or pass closely enough to solve for *n* tracer diffusion coefficients from 2(*n*-1) interdiffusion flux equations at the composition of intersection or rather at the closest compositions of body diagonal diffusion paths to each other. Hence, we now consider two diffusion couples at a time to calculate the tracer diffusion coefficients at the closest compositions to each other (which are near to the equi-atomic compositions). The thermodynamic parameters are extracted from TCHEA3 database of ThermoCalc [20], as given in Table 4. The estimated tracer diffusion coefficients for three pairwise combinations utilizing the three diffusion couples are given in Table 5. One may follow a least square method to solve 2(*n*-1) equations i.e., six equations in this quaternary system for calculation of four tracer diffusion coefficients. However, it may give more than one possible solution. Therefore, it is suggested first that four equations at a time should be considered first for one solution and consider the solution in that range from the least square method. It can be seen from Table 5 that the calculated tracer diffusion coefficient values are very consistent when estimated from different combination of diffusion couples. Moreover, the estimated values are found to have a very good match with the data estimated by the radiotracer method when extended to the temperature of measurement in this study [21]. This highlights the quality of solution obtained by following this method in multicomponent diffusion.

From the estimated tracer diffusion coefficients, we can now estimate the intrinsic diffusion coefficients utilizing Eq. 9 and interdiffusion coefficients utilizing Eq. 10. The estimated data are given in Table 6 and 7. It can be seen that the estimated intrinsic and interdiffusion coefficients from three combinations of body diagonal diffusion couples are very consistent.

One may also directly estimate the interdiffusion coefficients first from three diffusion profiles in a quaternary system and then solve for the estimation of the tracer diffusion coefficients by expressing the interdiffusion coefficients in terms of the tracer diffusion coefficients following Eq. 10. In this, nine interdiffusion coefficients are calculated considering the interdiffusion fluxes and composition gradients at the closest compositions. To locate the closest compositions considering three diffusion paths, a proximate composition range for each of the three profiles is selected by examining the 3D tetrahedral plot from different angular perspectives in Origin software. After converting these composition values to corresponding values of Cartesian co-ordinate, the mean position for a set of three points is calculated from

$$N\,(x, y, z) = \frac{1}{3}\sum_{i=1}^{3}\left(x_i, y_{i,} z_i\right) \qquad (8)$$

A simple iterative computer program that takes every possible combination of three points (each point from a different profile) within the selected range is utilized. Following, the program calculates the distances from the mean position to its corresponding set of points on the three different diffusion paths. Then the smallest value of the sum of these three distances is considered to identify the closest points between the three diffusion paths. These values are listed in Table 8. The mean position can be stated as the average composition at which the interdiffusion coefficients are estimated. The interdiffusion coefficients estimated using the flux and gradient values at these positions are listed in Table 9. These estimated values are found to be similar to those estimated from the tracer diffusion coefficients except for a few cross interdiffusion coefficients. Following, the tracer diffusion coefficients are estimated utilizing Eq. 10. Therefore, nine interdiffusion coefficients are solved following the least square method for four tracer diffusion coefficients. The estimated values are found to be $D_{Ni}^{*} = 5.3 \times 10^{-15}$, $D_{Co}^{*} = 5.6 \times 10^{-15}$, $D_{Fe}^{*} = 9.4 \times 10^{-15}$ and $D_{Cr}^{*} = 10.5 \times 10^{-15}$ $m^2/s$. Compared to the radiotracer measurements [21] and the tracer estimations considering only two profiles as done previously [Table 5], the Ni and Co tracer values from interdiffusion coefficients are slightly higher whereas the Fe and Cr tracer values show a good match. In this method, nine interdiffusion coefficients were first estimated from three closely passing diffusion paths and then a least squares approach was used to calculate the four tracer diffusion coefficients in the second step. This indirect approach has a higher likelihood of errors in the estimation compared to the previous method of using only two closely passing diffusion profiles that involves estimation of four tracers from six independent flux equations.

## 4. Conclusion

This study demonstrates the determination of the tracer coefficients and other diffusivities using only two diffusion profiles in a four component system that can also be extended to multicomponent systems with any number of components. Two diffusion profiles passing closely are sufficient for the estimation of all the tracer coefficients that can later be used to calculate the intrinsic coefficients as well. This is superior to the direct estimation of interdiffusion coefficients which requires higher number ($n$-1) of diffusion paths to intersect or pass closely and which do not give indications on the more fundamental intrinsic and tracer diffusion coefficients. For example, only after estimation of the tracer (or intrinsic) diffusion coefficients do we realize that $D_{Ni}^* \approx D_{Co}^* < D_{Fe}^* \leq D_{Cr}^*$, which is not apparent from the estimated interdiffusion coefficients.

It should be noted here that all the diffusion profiles produced following the concept of the body diagonal diffusion couple may not pass very closely to each other. In such a situation one may need to adjust the end member compositions again with the aim of bringing the diffusion paths closer to each other. This exercise can be avoided when the method demonstrated in this article is followed if at least two diffusion paths pass closely among the ($n$-1) diffusion profiles, thus reducing the effort significantly. Further, it is not necessary that one has to follow the body diagonal diffusion couple method only for such analysis. One may consider any two profiles passing closely irrespective of the design methodology followed for producing the diffusion couples.

Until recently, the estimation of the diffusion coefficients in an inhomogeneous material following the diffusion couple method was considered impossible in a system with more than three components. We have now different methods such as pseudo-binary [2, 4, 22], pseudo-ternary [3, 4, 22, 23] and body-diagonal diffusion couple methods [5] for estimation of the diffusion coefficients. All these methods have different advantages and limitations. One may select one particular or a combination of these methods to produce reliable diffusion coefficients of all the components since every method may not be suitable in all the systems (unlike the NiCoFeCr system considered in this study) depending on the composition range or difficulties of producing the diffusion couples in a certain way fulfilling the conditions of such diffusion profiles. The reliability of data generated can be realized from the fact that the tracer diffusion coefficients estimated following these three different methods (estimated following the pseudo-binary and pseudo-ternary methods in previous studies [4, 22, 24] and body diagonal diffusion couple method in this study) fall in a close range showing a very good match with the tracer diffusion coefficients estimated following the radiotracer method [21], as shown in Fig. 3. The average and data range from different methods match nicely with the average and range of data estimated following the radio-tracer method. One can now extend such analysis to various Al, Si, Pt etc. containing alloys in which the radio-tracer method cannot be followed due to short half-life or costly radioisotopes and even without being restricted by constraints for maintaining such facility because of stringent safety regulations, which is now practiced only in very few groups around the world for these types of studies.

**Acknowledgement**

AP dedicates this article to the memory of Late Prof. John Morral for his help, guidance, and discussion on numerous occasions at the initial stage of his professional career. This work extends and strengthens the body diagonal diffusion couple method in multicomponent system proposed by him. We also acknowledge the help from Prof. Sergiy Diviniski for plotting the error range of tracer diffusion coefficients estimated following the radiotracer method. We also acknowledge the help from Dr. N. Esakkiraja at the initial stage of this study.

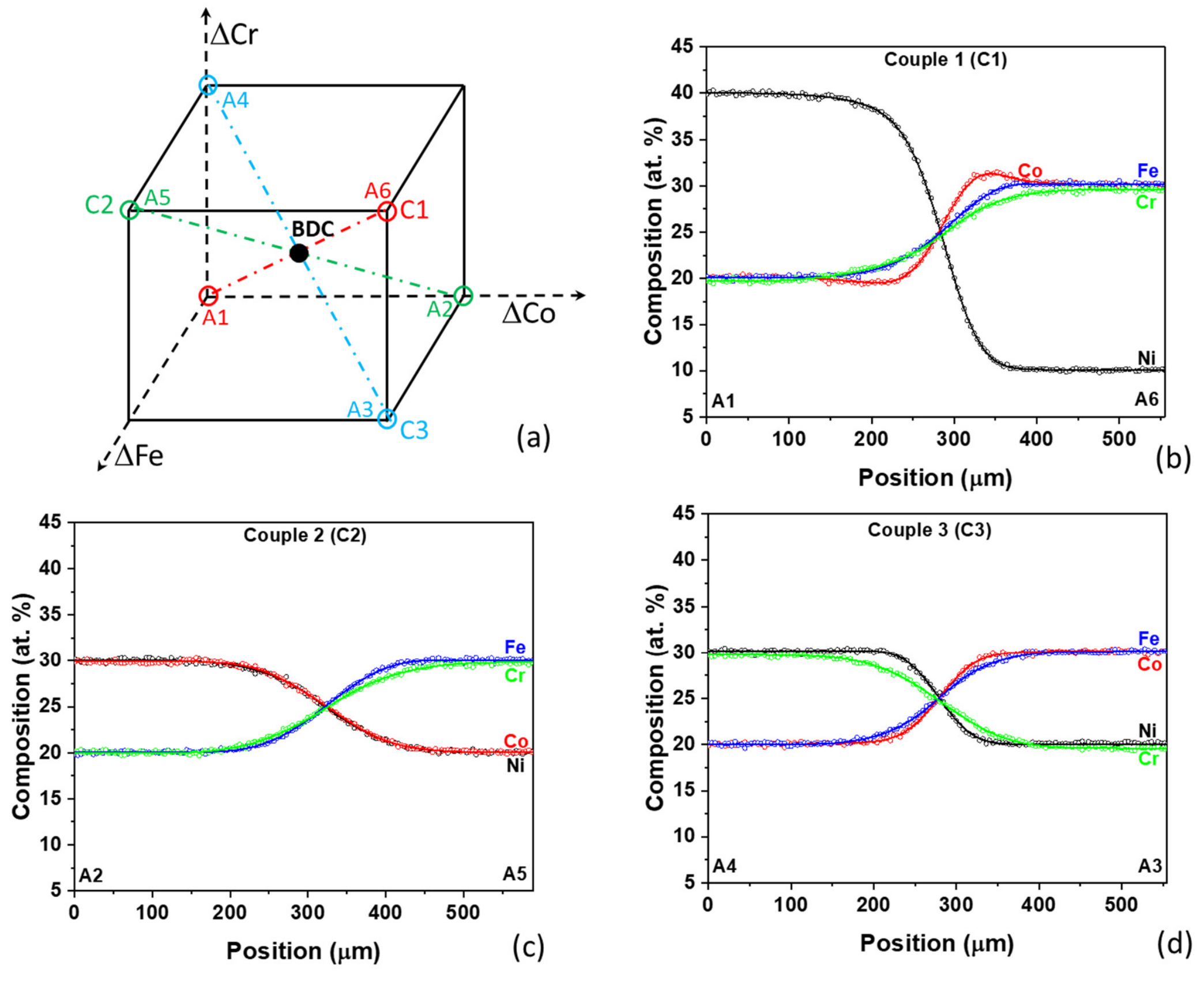

Fig. 1 (a) The end-member alloys A1-A6 used for producing three body diagonal diffusion couples C1-C3 (see Table 1). BDC is the body center equiatomic composition. (b-d) The composition profiles of the diffusion couples. DCo, DFe, DCr are considered as 10 at.%.

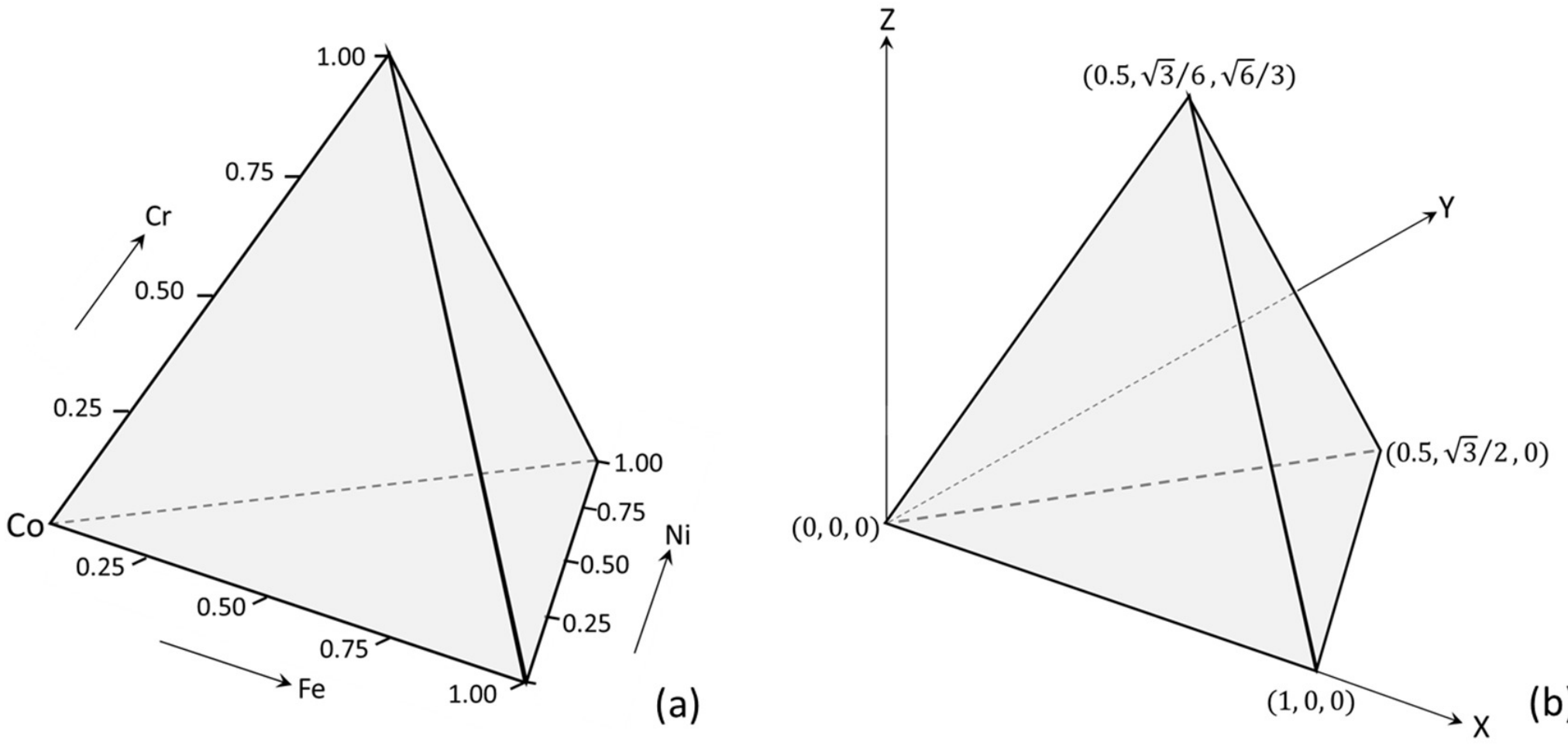

Fig. 2: Representation of tetrahedral phase diagram in barycentric coordinates and its equivalent cartesian position.

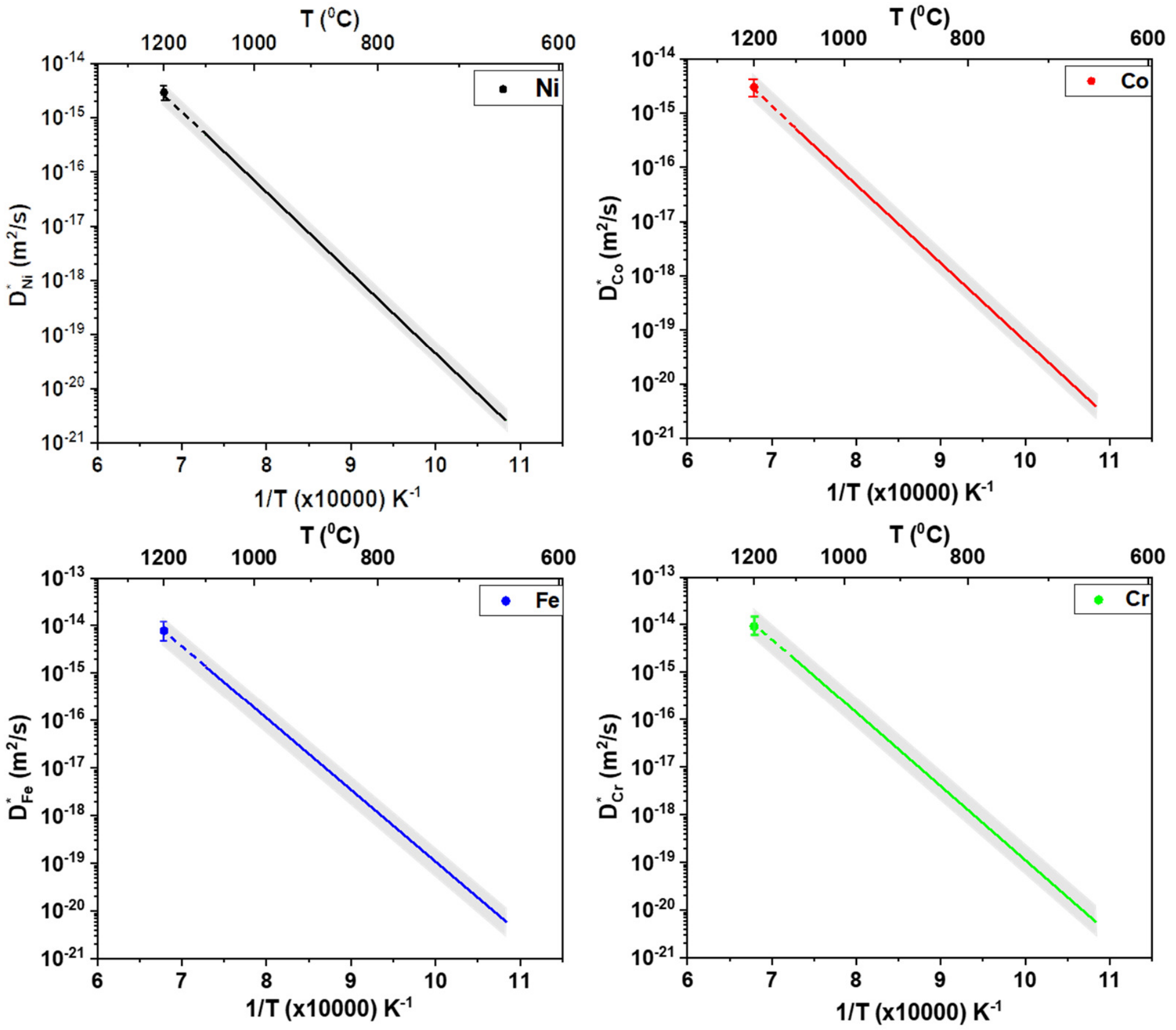

Fig. 3: Comparison of tracer diffusion coefficients estimated following the radio-tracer method [21] with the data estimated following the pseudo-binary, pseudo-ternary and body-diagonal diffusion couple methods.

| | | Co (at. %) planned (actual) | Fe (at. %) planned (actual) | Cr (at. %) planned (actual) | Ni (at. %) planned (actual) |
|---|---|---|---|---|---|
| C1 | A1 | 20 (20.2) | 20(20.1) | 20(19.7) | 40(40) |
| | A6 | 30 (30.1) | 30(30.2) | 30(29.7) | 10(10) |
| C2 | A2 | 30 (30) | 20(20.1) | 20(19.9) | 30(30) |
| | A5 | 20 (20) | 30(30.1) | 30(29.8) | 20(20.1) |
| C3 | A3 | 30 (30.2) | 30(30.1) | 20(19.6) | 20(20.1) |
| | A4 | 20(20) | 20(20) | 30(29.8) | 30(30.2) |

Table 1: The planned and actual compositions of end-member compositions for diffusion couples. Six alloys A1-A6 are coupled to produce three diffusion couples C1-C3, which are shown in Fig. 1.

| Tetrahedral space | Cartesian space position $(x, y, z)$ |
|---|---|
| Pure Ni ($N_{Ni}$ = 1) | $(0.5, \sqrt[2]{3}/2, 0)$ |
| Pure Co ($N_{Co}$ = 1) | $(0, 0, 0)$ |
| Pure Fe ($N_{Fe}$ = 1) | $(1, 0, 0)$ |
| Pure Cr ($N_{Cr}$ = 1) | $(0.5, \sqrt[2]{3}/6, \sqrt[2]{6}/3)$ |

Table 2: Cartesian space positions corresponding to pure components at vertices of a tetrahedral phase diagram

| | Composition (at. frac.) | | | | Cartesian position | | | Distance to mean |
|---|---|---|---|---|---|---|---|---|
| | Ni | Co | Fe | Cr | $x$ | $y$ | $z$ | |
| C1 | 0.253 | 0.253 | 0.249 | 0.246 | 0.498 | 0.290 | 0.201 | 0.0005 |
| C2 | 0.252 | 0.253 | 0.248 | 0.247 | 0.498 | 0.290 | 0.202 | |
| Mean | 0.253 | 0.253 | 0.249 | 0.247 | 0.498 | 0.290 | 0.202 | |

| | Composition (at. frac.) | | | | Cartesian position | | | Distance to mean |
|---|---|---|---|---|---|---|---|---|
| | Ni | Co | Fe | Cr | $x$ | $y$ | $z$ | |
| C2 | 0.250 | 0.251 | 0.250 | 0.248 | 0.499 | 0.288 | 0.202 | 0.0003 |
| C3 | 0.251 | 0.251 | 0.251 | 0.248 | 0.500 | 0.289 | 0.202 | |
| Mean | 0.251 | 0.251 | 0.251 | 0.248 | 0.500 | 0.289 | 0.202 | |

| | Composition (at. frac.) | | | | Cartesian position | | | Distance to mean |
|---|---|---|---|---|---|---|---|---|
| | Ni | Co | Fe | Cr | $x$ | $y$ | $z$ | |
| C1 | 0.249 | 0.255 | 0.250 | 0.246 | 0.498 | 0.287 | 0.201 | 0.0011 |
| C3 | 0.248 | 0.253 | 0.252 | 0.247 | 0.499 | 0.286 | 0.202 | |
| Mean | 0.249 | 0.254 | 0.251 | 0.247 | 0.499 | 0.287 | 0.202 | |

Table 3: Nearest points for each pair of profiles from C1, C2 and C3 and equivalent distance between points in cartesian coordinates.

| $\phi_{CrNi}^{Cr}$ | $\phi_{CrCo}^{Cr}$ | $\phi_{CrFe}^{Cr}$ |
|---|---|---|
| -1.35 | -1.41 | -1.16 |
| $\phi_{NiNi}^{Cr}$ | $\phi_{NiCo}^{Cr}$ | $\phi_{NiFe}^{Cr}$ |
| 1.14 | 0.25 | 0.08 |
| $\phi_{CoNi}^{Cr}$ | $\phi_{CoCo}^{Cr}$ | $\phi_{CoFe}^{Cr}$ |
| 0.31 | 1.31 | 0.08 |
| $\phi_{FeNi}^{Cr}$ | $\phi_{FeCo}^{Cr}$ | $\phi_{FeFe}^{Cr}$ |
| -0.11 | -0.16 | 0.99 |

Table 4: Thermodynamic parameters extracted from TCHEA3 database of ThermoCalc.

| | $D^*_{Ni}$ (×10$^{-15}$ m$^2$/s) | $D^*_{Co}$ (×10$^{-15}$ m$^2$/s) | $D^*_{Fe}$ (×10$^{-15}$ m$^2$/s) | $D^*_{Cr}$ (×10$^{-15}$ m$^2$/s) |
|---|---|---|---|---|
| C1 - C2 | 3.3 | 4.1 | 12.7 | 14.3 |
| C2 - C3 | 3.8 | 4.6 | 10.6 | 12 |
| C1 - C3 | 3.8 | 3.2 | 9.1 | 10.6 |
| Average | 3.6±0.3 | 4±0.7 | 10.8±1.8 | 12.3±2 |

Table 5: Estimated tracer diffusion coefficients from body-diagonal diffusion couples

| | $D^{Cr}_{NiNi}$ (×10$^{-15}$ m$^2$/s) | $D^{Cr}_{NiCo}$ (×10$^{-15}$ m$^2$/s) | $D^{Cr}_{NiFe}$ (×10$^{-15}$ m$^2$/s) | $D^{Cr}_{CoNi}$ (×10$^{-15}$ m$^2$/s) | $D^{Cr}_{CoCo}$ (×10$^{-15}$ m$^2$/s) | $D^{Cr}_{CoFe}$ (×10$^{-15}$ m$^2$/s) | $D^{Cr}_{FeNi}$ (×10$^{-15}$ m$^2$/s) | $D^{Cr}_{FeCo}$ (×10$^{-15}$ m$^2$/s) | $D^{Cr}_{FeFe}$ (×10$^{-15}$ m$^2$/s) | $D^{Cr}_{CrNi}$ (×10$^{-15}$ m$^2$/s) | $D^{Cr}_{CrCo}$ (×10$^{-15}$ m$^2$/s) | $D^{Cr}_{CrFe}$ (×10$^{-15}$ m$^2$/s) |
|---|---|---|---|---|---|---|---|---|---|---|---|---|
| C1 – C2 | 3.76 | 0.82 | 0.25 | 1.27 | 5.30 | 0.34 | -1.45 | -2.06 | 12.58 | -19.07 | -19.63 | -16.35 |
| C1 – C3 | 4.33 | 0.95 | 0.29 | 1.01 | 4.23 | 0.27 | -1.04 | -1.48 | 9.04 | -14.14 | -14.55 | -12.12 |
| C2 – C3 | 4.34 | 0.95 | 0.29 | 1.43 | 5.97 | 0.38 | -1.21 | -1.73 | 10.53 | -15.99 | -16.46 | -13.71 |
| Average | 4.14 | 0.91 | 0.28 | 1.24 | 5.17 | 0.33 | -1.23 | -1.76 | 10.72 | -16.40 | -16.88 | -14.06 |

Table 6: Estimated intrinsic diffusion coefficients from the tracer diffusion coefficients.

| | $\tilde{D}_{NiNi}^{Cr}$ (×10$^{-15}$ m$^2$/s) | $\tilde{D}_{NiCo}^{Cr}$ (×10$^{-15}$ m$^2$/s) | $\tilde{D}_{NiFe}^{Cr}$ (×10$^{-15}$ m$^2$/s) | $\tilde{D}_{CoNi}^{Cr}$ (×10$^{-15}$ m$^2$/s) | $\tilde{D}_{CoCo}^{Cr}$ (×10$^{-15}$ m$^2$/s) | $\tilde{D}_{CoFe}^{Cr}$ (×10$^{-15}$ m$^2$/s) | $\tilde{D}_{FeNi}^{Cr}$ (×10$^{-15}$ m$^2$/s) | $\tilde{D}_{FeCo}^{Cr}$ (×10$^{-15}$ m$^2$/s) | $\tilde{D}_{FeFe}^{Cr}$ (×10$^{-15}$ m$^2$/s) |
|---|---|---|---|---|---|---|---|---|---|
| C1-C2 | 7.64 | 4.72 | 1.05 | 5.18 | 9.23 | 1.14 | 2.42 | 1.83 | 13.38 |
| C1-C3 | 6.79 | 3.67 | 0.92 | 3.50 | 6.97 | 0.91 | 1.42 | 1.23 | 9.67 |
| C2-C3 | 7.20 | 3.77 | 0.92 | 4.32 | 8.81 | 1.02 | 1.65 | 1.09 | 11.16 |
| Average | 7.21 | 4.05 | 0.96 | 4.33 | 8.34 | 1.02 | 1.83 | 1.38 | 11.40 |

Table 7: Estimated interdiffusion coefficients from the tracer diffusion coefficients.

| | Composition (at. frac.) | | | | Cartesian position | | | Distance to mean |
|---|---|---|---|---|---|---|---|---|
| | Ni | Co | Fe | Cr | $x$ | $y$ | $z$ | |
| C1 | 0.252 | 0.253 | 0.249 | 0.246 | 0.499 | 0.289 | 0.201 | 0.0013 |
| C2 | 0.251 | 0.252 | 0.249 | 0.247 | 0.499 | 0.289 | 0.202 | 0.0005 |
| C3 | 0.251 | 0.251 | 0.251 | 0.248 | 0.500 | 0.289 | 0.202 | 0.0012 |
| Mean | 0.251 | 0.252 | 0.250 | 0.247 | 0.499 | 0.289 | 0.202 | |

Table 8: Nearest three points for C1, C2 and C3, their mean position and distance from each position to mean position

| Interdiffusion Coefficients (x $10^{-15}$) $m^2/s$ | | |
|---|---|---|
| $\widetilde{D}_{NiNi}^{Cr}$ | $\widetilde{D}_{NiCo}^{Cr}$ | $\widetilde{D}_{NiFe}^{Cr}$ |
| 7.5 | 5.3 | 0.05 |
| $\widetilde{D}_{CoNi}^{Cr}$ | $\widetilde{D}_{CoCo}^{Cr}$ | $\widetilde{D}_{CoFe}^{Cr}$ |
| 4.9 | 9 | 1 |
| $\widetilde{D}_{FeNi}^{Cr}$ | $\widetilde{D}_{FeCo}^{Cr}$ | $\widetilde{D}_{FeFe}^{Cr}$ |
| -0.01 | -0.8 | 9.15 |

Table 9: A direct estimation of the interdiffusion coefficients from three diffusion profiles